\documentclass[runningheads]{llncs}
\usepackage[T1]{fontenc}
\usepackage{graphicx}
\usepackage[colorlinks=true, allcolors=blue]{hyperref}
\usepackage{color}

\usepackage{multirow}
\usepackage{amsmath}
\usepackage{siunitx}
\usepackage{booktabs}
\usepackage[flushleft]{threeparttable}
\usepackage{graphicx}

\begin{document}
\title{Copy Number Variation Informs fMRI-based Prediction of Autism Spectrum Disorder}
\titlerunning{Copy Number Variation Informs fMRI-based Prediction of ASD}
%
\author{Nicha C. Dvornek\inst{1,2}
\and Catherine Sullivan\inst{3}
\and James S. Duncan\inst{1,2}
\and Abha R. Gupta\inst{3}}
\authorrunning{N. C. Dvornek et al.}
%
\institute{Department of Radiology \& Biomedical Imaging, Yale School of Medicine, New Haven, CT 06510, USA \and
Department of Biomedical Engineering, Yale University, New Haven, CT 06511, USA \and
Department of Pediatrics, Yale School of Medicine, New Haven, CT 06510, USA 
\email{\{nicha.dvornek, catherine.sullivan, james.duncan, abha.gupta\}@yale.edu}}
\maketitle              
\begin{abstract}
The multifactorial etiology of autism spectrum disorder (ASD) suggests that its study would benefit greatly from multimodal approaches that combine data from widely varying platforms, e.g., neuroimaging, genetics, and clinical characterization. Prior neuroimaging-genetic analyses often apply naive feature concatenation approaches in data-driven work or use the findings from one modality to guide posthoc analysis of another, missing the opportunity to analyze the paired multimodal data in a truly unified approach. In this paper, we develop a more integrative model for combining genetic, demographic, and neuroimaging data. Inspired by the influence of genotype on phenotype, we propose using an attention-based approach where the genetic data guides attention to neuroimaging features of importance for model prediction. The genetic data is derived from copy number variation parameters, while the neuroimaging data is from functional magnetic resonance imaging. We evaluate the proposed approach on ASD classification and severity prediction tasks, using a sex-balanced dataset of 228 ASD and typically developing subjects in a 10-fold cross-validation framework. We demonstrate that our attention-based model combining genetic information, demographic data, and functional magnetic resonance imaging results in superior prediction performance compared to other multimodal approaches.

\end{abstract}
\section{Introduction}
Autism spectrum disorder (ASD) is characterized by impaired communication and social skills, and restricted, repetitive, and stereotyped behaviors that result in significant disability \cite{DSMV}. ASD refers to a spectrum of disorders due to its heterogeneity, with multiple etiologies, sub-types, and developmental trajectories \cite{masi2017overview}, resulting in diverse clinical presentation of symptoms and severity. Two major factors contributing to the heterogeneity of ASD include \textit{genetic variability} and \textit{sex} \cite{masi2017overview}. Research aimed at uncovering the pathophysiology of ASD and its heterogeneous presentations is critical to reduce disparities in early diagnosis and develop personalized targeted treatments. 

A popular data-driven approach for discovering biomarkers for ASD is to first build a classification model that can distinguish ASD vs.\ typically developing (TD) individuals. Prior work generally focuses on unimodal data, e.g., functional magnetic resonance imaging (fMRI), structural MRI, genetics, or behavioral or developmental scores alone \cite{hyde2019applications}. However, given the multifactorial etiology of ASD \cite{lord2020autism}, a unified multimodal approach should improve model classification performance. Furthermore, convergence between different modality datasets on where in the brain ASD may arise and which brain regions correlate with different clinical measures would provide greater confidence in the results.

Prior multimodal methods combining genetic and neuroimaging data often naively concatenate the multimodal features and use them as inputs in a machine learning algorithm \cite{bi2021novel,yang2010hybrid,heinrich2016prediction}.  We aim to integrate the multimodal data in a more informative model design. Furthermore, such concatenation approaches may suffer from differing scales of the multimodal data, which will artificially give one modality greater importance. Another major direction is to use the findings from one modality to guide posthoc analysis of another \cite{jack2021neurogenetic,antonucci2019thalamic,vertes2016gene}, missing the opportunity to analyze the paired multimodal data in a truly unified approach. While phenotypic data has been combined with neuroimaging data in multiple ASD studies using thoughtful model designs \cite{parisot2017spectral,dvornek2018learning,dvornek2018combining}, such integration between genetic and neuroimaging data has not been explored.

Here, we propose to improve characterization of the neurobiology of ASD by developing an integrated neuroimaging-genetic deep learning model to accurately classify ASD vs.\ TD individuals and predict ASD severity. As individual genetic differences will influence neuroimaging phenotypes, we propose using an attention-based approach where genetic variables inform what neuroimaging features should be attended to for model prediction. We assess the performance of the proposed approach on ASD classification and severity prediction tasks, using a sex-balanced dataset of 228 ASD and TD subjects in a 10-fold cross-validation framework. We demonstrate superior performance to other methods of combining multimodal data. 

\section{Methods}
\subsection{Dataset and preprocessing}
The dataset includes a sex-balanced cohort of 228 youth (age range: $8.0 - 17.9$ years), 114 with ASD (59 female, 55 male) and 114 TD controls (58 female, 56 male), available from NIH NDA collection 2021. Data types utilized include clinical measures, genome-wide genotyping, and neuroimaging. 

Clinical measures utilized include age at time of scan, sex, ASD diagnosis, and Social Responsiveness Scale-2 (SRS, range: $1-162$) \cite{constantino2012social}. SRS is a measure of severity of social impairment in ASD, but was assessed on both ASD and TD subjects. 

Genome-wide genotyping data was generated using the HumanOmni 2.5M BeadChip (Illumina). This data was processed and analyzed for rare (>50\% of copy number variants (CNV) at <1\% frequency in the Database of Genomic Variants) genic CNVs \cite{jack2021neurogenetic}. CNV parameters of number of CNVs (range: $0 - 10)$ and summed total length of all CNVs (range: $0 - 4050194$ bp) were used as model inputs. 

Structural and fMRI data were acquired under several different tasks. We utilized the Biopoint task-fMRI acquisition, in which subjects viewed coherent and scrambled biological point-light animations in alternating blocks (24s per block; 154 volumes; TR = 2000 ms; TE = 30 ms; flip angle = 90°; FOV = 192 mm; image matrix = 64 mm$^2$; voxel size = 3 mm × 3 mm × 4 mm; 34 slices). Prior Biopoint studies have identified dysfunction in biological motion processing as reflecting key neural signatures of ASD and as a neuroendophenotype of genetic risk in unaffected siblings \cite{kaiser2010neural,bjornsdotter2016evaluation}. fMRI data was preprocessed in FSL \cite{jenkinson2012fsl}, including MCFLIRT motion correction, interleaved slice timing correction, BET brain extraction, spatial smoothing (FWHM 5mm), high-pass temporal filtering, and registration to Montreal Neurological Institute space. We parcellated the brain into 116 regions of interest (ROIs) with the AAL atlas \cite{tzourio2002automated}. Standardized mean time-series from each ROI were used as model inputs.

\subsection{Network architecture}

\begin{figure}[t]
\centering
\includegraphics[width=0.8\textwidth,keepaspectratio]{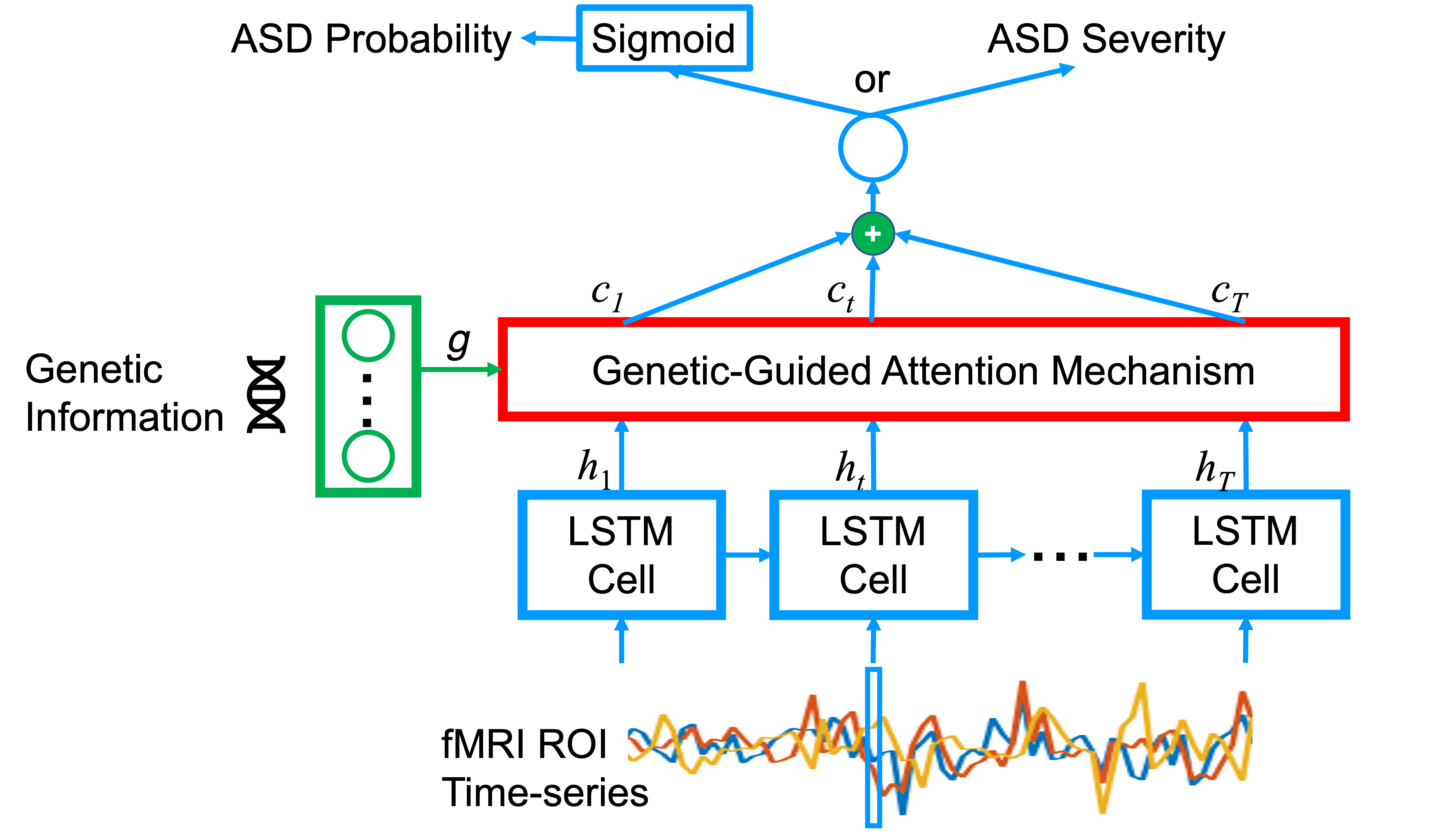}
\caption{The proposed integrated neuroimaging-
genetic RNN model. The genetic information will be
used to focus attention to different fMRI features for the
prediction task.} 
\label{network}
\end{figure}

The structure of the proposed genetic-neuroimaging model is shown in Fig.\ \ref{network}. The fMRI time-series from predefined ROIs is first input to a long short-term memory (LSTM) layer \cite{hochreiter1997long} to encode fMRI information \cite{dvornek2017identifying}. Then, inspired by the influence of genotype on phenotype, we propose using a generalized attention mechanism \cite{vaswani2017attention} to steer focus to different fMRI features according to the genetic information. We utilize the genetic data of CNV number and total CNV size derived from genome-wide genotyping, since larger CNV size has been associated with deleteriousness in previous ASD genetics studies \cite{jacquemont2014higher,pinto2014convergence}.

The generalized attention mechanism can be defined as a mapping between a query and key-value pair to the “context”. Here, we define the query by the genetic and demographic data $g\in{R^G}$, and the key and value are defined by the encoded fMRI data, which are the outputs of the LSTM $h_t\in{R^L}$. Applying scaled dot product attention \cite{vaswani2017attention}, the context is computed as
\begin{equation}
c_t=att(g,{h_t})=softmax\left[\frac{\left(W_{q}g\right)^T\left(W_{k}h_t\right)}{\sqrt{M}}\right]W_{v}h_t,
\end{equation}
where $softmax(a_t)$ normalizes $a_t$ such that $\sum_{t=1}^{T}{a_t}=1$, $W_q$ encodes the genetic and demographic information $g$ into the query, $W_k$ and $W_v$ encode fMRI-based $h_t$ into the key and value, respectively, and $M$ denotes the dimension of the encoded space. Multiple attention mappings with different encodings $W$ could be learned to diversify ways in which genetic information modulates the fMRI information. We then summarize the context across the $T$ timesteps $\sum_{t=1}^{T}{c_t}$ and apply a fully connected layer to predict the output from the summary $M$ context features. The ASD classification model ends with a sigmoid activation function to produce the probability of the ASD class, while the ASD severity regression model ends with the output of the fully connected layer.

\subsection{Experimental Settings}

To assess the effectiveness of our multimodal attention-based approach (\textit{Att}), we compared to 4 other methods:

\noindent1) \textit{Base}: A basic LSTM model using only the fMRI time-series data \cite{dvornek2017identifying}.

\noindent2) \textit{Concat}: Concatenation of the genetic and demographic data with the fMRI time-series data. The genetic and demographic values are repeated across time and concatenated as additional features to the fMRI data. The data are then input to the basic LSTM model.

\noindent3) \textit{Fusion}: Fusion of the genetic and demographic information with the prediction from the LSTM processed fMRI data \cite{dvornek2018combining}. The genetic and demographic information is combined with the LSTM block using a fully connected layer, followed by a layer with a single node for the task prediction.

\noindent4) \textit{LSTMinit}: The baseline LSTM model with initialization of hidden states conditioned on the genetic and demographic variables \cite{dvornek2018learning}. The LSTM initial state vectors are learned using a single fully connected layer with size $L$.

We also perform an ablation study to assess the utility of including both genetic and demographic information to guide the attention to fMRI features in the proposed full model. We thus trained reduced models that included only demographic (\textit{Att-demo}) or genetic (\textit{Att-gene}) information alone to evaluate their value. 

All models were implemented in Python using Keras \cite{chollet2015keras} and Tensorflow \cite{tensorflow2015-whitepaper} libraries.  The feature dimension for the LSTM output was $L=16$, and the feature dimension for the attention embedding was set to $M=8$. fMRI time-series for each ROI were standardized to have mean 0 and standard deviation 1, and genetic and demographic inputs were normalized to the range [-1,1]. SRS raw scores were normalized to the range [0,1]. We applied randomly shifted windowed samples of the fMRI time-series data as data augmentation \cite{dvornek2017identifying}, sampling 10 random windows of size $T=48$ (representing the time for 2 blocks each of scrambled and biological motion) per subject every epoch. We used the Adam optimizer \cite{kingma2017adam} ($lr = 0.001$) to train the models for up to 50 epochs. Classification models were trained to optimize binary cross-entropy loss and severity regression models were trained to optimize mean squared error loss.

To evaluate model performance, we performed 10-fold cross-validation of subjects, using stratified sampling of ASD status, with 80\% subjects for training, 10\% for validation, and 10\% for testing in each partition. Validation loss was used to determine the stopping epoch for model selection. For classification models, we computed classification accuracy, sensitivity, specificity, and area under the receiver operating characteristic curve (AUC) for each fold. We also computed the overall AUC based on aggregating the test predictions from all folds. For regression models, we computed the mean squared error (MSE), maximum of the mean squared error, and Pearson correlation between true and predicted test outputs for each fold. Because Pearson correlation can vary largely when there are small perturbations in a small test dataset, we also computed the overall Pearson correlation based on aggregating the test predictions from all folds. Significant differences between models were assessed using paired two-tailed t-tests to compare matched cross-validation folds ($\alpha$ = 0.05).  

\section{Results}
\subsection{ASD Classification}

\subsubsection{Classification performance}
\begin{figure}[h]
\centering
\includegraphics[width=0.65\textwidth,keepaspectratio]{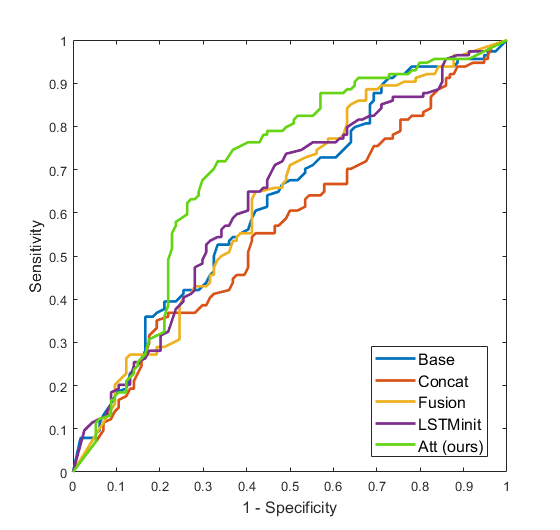}
\caption{Receiver operating characteristic curves from ASD classification models.}
\label{fig:rocs}
\end{figure}

ASD classification model results are summarized in Table \ref{tab:class_results}, and receiver operating characteristic curves are plotted in Fig.\ \ref{fig:rocs}. Traditional concatenation of multimodal inputs performed worse than the fMRI-only base model. Fusing multimodal feature in later layers produced the highest sensitivity, but was not significantly different from the base or our attention-based model results. Initializing the LSTM with genetic and demographic information performed similarly to the fMRI-only base model. Our attention-based approach using the genetic information to inform the attending of important fMRI features resulted in the highest classification accuracy, specificity, and AUC. Furthermore, our attention-based model is the only multimodal approach to perform significantly better than the fMRI-only base model based on the accuracy metric. Moreover, our approach significantly outperformed the other multimodal models as measured by accuracy, specificity, and AUC. As seen in Fig.\ \ref{fig:rocs}, the receiver operating characteristic curve for our attention model lies above the other models for much of the plot.   

When training the attention-based model with only demographic or genetic data for guidance, the models still produced generally better results than the non-attention models, but with a small performance drop compared to the full model utilizing both demographic and genetic information. Using genetic information alone resulted in significantly worse specificity compared to the full model. Again, only the full model utilizing both genetic and demographic data performed significantly better than the unimodal fMRI base model (as measured by accuracy), suggesting the importance of including all modalities in computing the attention scores. 

\begin{table}[t]
\centering
\caption{ASD classification performance (mean ± standard deviation). Best results marked in bold.}
\label{tab:class_results}
\begin{threeparttable}
\begin{tabular}{l|c|c|c|c|c}
\toprule
Model & Accuracy\enspace & Sensitivity  & Specificity & AUC & Overall AUC\\
\hline\hline
Base & 0.59 ± 0.13$^{*}$ & 0.67 ± 0.19 & 0.52 ± 0.22\enspace & 0.63 ± 0.14 & 0.62 \\
\hline
Concat & 0.54 ± 0.07$^{*}$ & 0.64 ± 0.14 & 0.44 ± 0.15$^{*}$ & \enspace\enspace0.56 ± 0.12$^{*\dag}$ & 0.56\\
\hline
Fusion & 0.60 ± 0.12$^{*}$ & \textbf{0.73 ± 0.11} & 0.46 ± 0.20$^{*}$ & 0.63 ± 0.13 & 0.62\\
\hline
LSTMinit & 0.62 ± 0.11\enspace & 0.69 ± 0.12 & 0.54 ± 0.12$^{*}$ & 0.64 ± 0.14 & 0.63\\
\hline\hline
Att-demo & 0.68 ± 0.06\enspace & 0.67 ± 0.13 & 0.68 ± 0.11\enspace & \textbf{0.70 ± 0.08} & 0.69\\ 
\hline
Att-gene & 0.67 ± 0.06\enspace & 0.68 ± 0.11 & 0.65 ± 0.13$^*$ & 0.69 ± 0.08 & 0.69\\ 
\hline
Att (ours)\enspace & \textbf{0.69 ± 0.06$^\dag$} & 0.68 ± 0.13 & \textbf{0.69 ± 0.11}\enspace & \textbf{0.70 ± 0.08} & \textbf{0.70}\\ 
\bottomrule
\end{tabular}
\begin{tablenotes}
\scriptsize{
\item$^{*}$ Significant difference compared to our proposed approach Att ($p < 0.05$, paired two-tailed t-test)
\item$^\dag$ Significant difference compared to fMRI-only Base model ($p < 0.05$, paired two-tailed t-test)}
\end{tablenotes}
\end{threeparttable}
\end{table}

\subsubsection{Model analysis}

\begin{figure}[t]
\begin{minipage}[b]{0.7\columnwidth}
\centering
\includegraphics[width=0.95\textwidth,keepaspectratio]{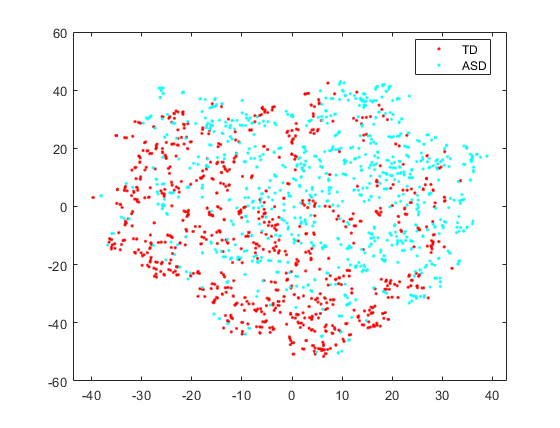}
\caption{tSNE visualization of the summary context vector for each sample.}
\label{fig:tsne}
\end{minipage}
\hfill
\begin{minipage}[b]{0.25\columnwidth}
\centering
\includegraphics[width=\textwidth,keepaspectratio]{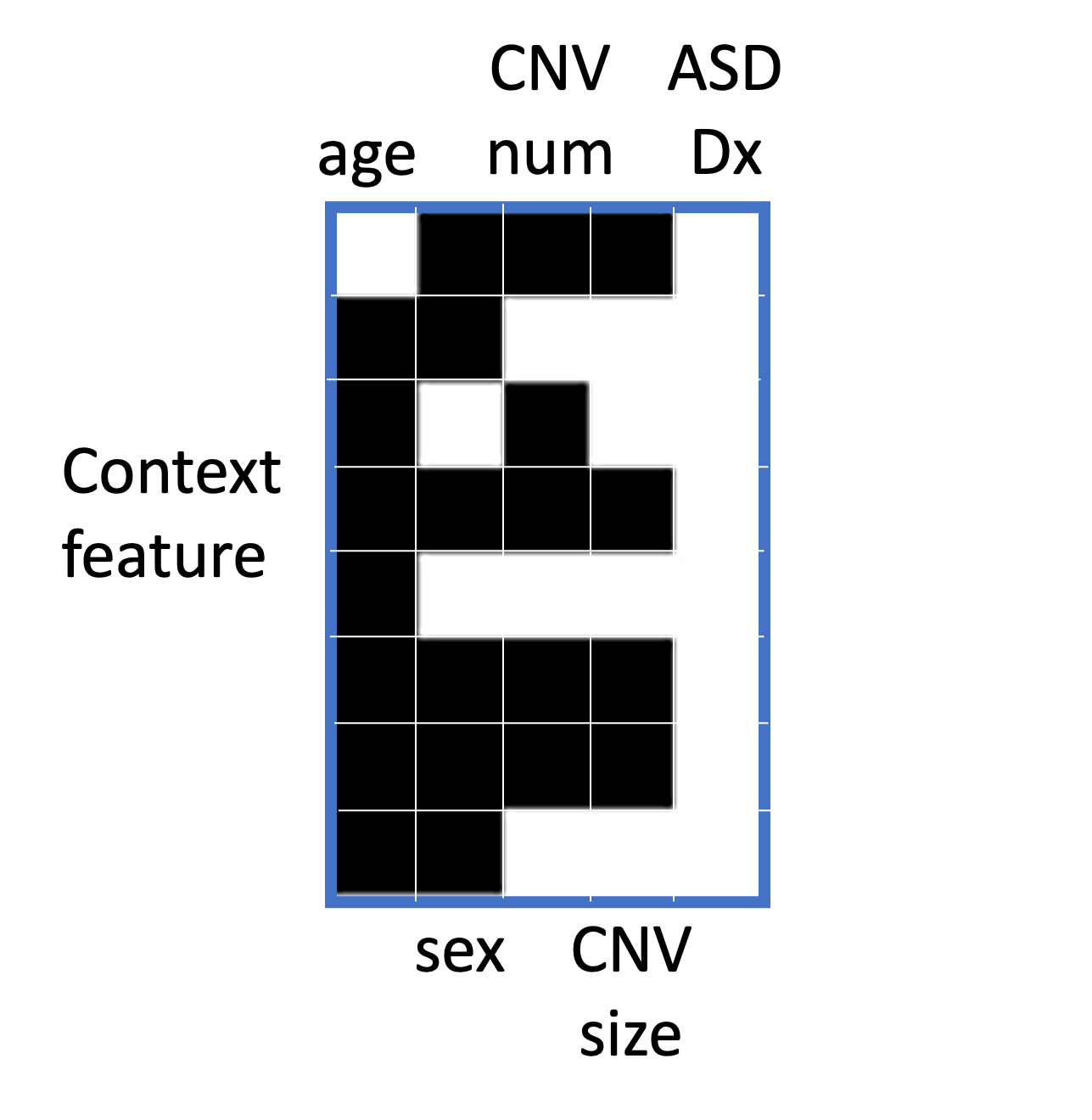}
\caption{Context features are associated with different genetic and demographic features.}
\label{fig:corrs}
\end{minipage}
\end{figure}

We investigated our attention-based model's representation of multimodal information via the summary context vector. First, we performed a tSNE visualization of the context for one trained model (Fig.\ \ref{fig:tsne}), where TD subjects are plotted in red and ASD subjects are plotted in blue. We see that there is clustering of the TD and ASD samples.

We then investigate the impact of each genetic and demographic variable on each context feature. We compute the Pearson correlation between a genetic or demographic variable $g(i)$ and the ASD diagnosis and each feature of the context representation $c(j)$ across all samples in a representative fold. We visualize the input and output features with significant correlations ($p < 0.05$) for each context feature by white boxes in Fig. \ref{fig:corrs}. As expected, every context feature is correlated with the ASD diagnosis. Furthermore, we see that different context features are guided by different input features. For example, the context feature corresponding to row 1 has only 1 feature, i.e., age, associated with it, while the second context feature is associated with both genetic variables. Our attention-based approach allows for different modalities of inputs to be integrated in a dynamic manner to produce more informative, individualized representations of the input data, which lends to the improved classification performance.

\subsection{ASD Severity Regression}
ASD severity regression results are summarized in Table \ref{tab:reg_results}. The fMRI-only base model was able to achieve a significant overall correlation. Concatenating genetic and demographic features with fMRI nominally improved the prediction performance. Fusing the genetic and demographic features with fMRI features significantly reduced the MSE compared to the fMRI-only base model and produced the lowest maximum errors; however, the correlation between predicted and true scores was greatly reduced and no longer significant. The model initializing the LSTM with genetic and demographic variables performed slightly worse than the base fMRI model. Our attention-based approach resulted in the lowest MSE and highest correlation.

\begin{table}[t]
\centering
\caption{ASD severity prediction performance (normalized SRS mean ± standard deviation). Best results marked in bold.}
\label{tab:reg_results}
\begin{threeparttable}
\begin{tabular}{l|c|c|c|c}
\toprule
Model & MSE\enspace & Maximum  & Correlation & Overall Correlation\\
\hline\hline
Base & 0.084 ± 0.014$^{*}$ & 0.30 ± 0.09 & 0.16 ± 0.19 & 0.15$^\star$ \\
\hline
Concat & 0.080 ± 0.016\enspace & 0.29 ± 0.06 & 0.17 ± 0.21 & 0.16$^\star$\\
\hline
Fusion & 0.075 ± 0.016$^\dag$ & \textbf{0.26 ± 0.08} & 0.08 ± 0.12 & 0.08\enspace\\
\hline
LSTMinit & 0.087 ± 0.011\enspace & 0.28 ± 0.08 & 0.13 ± 0.28 & 0.12\enspace\\
\hline\hline
Att-demo$\quad$ & \textbf{0.074 ± 0.019}$^\dag$ & 0.29 ± 0.09 & \textbf{0.19 ± 0.11} & \textbf{0.18}$^{\star\star}$\\ 
\hline
Att-gene$\quad$ & \textbf{0.074 ± 0.019$^\dag$} & 0.29 ± 0.08 & \textbf{0.19 ± 0.10} & 0.16$^\star$\\ 

\hline
Att (ours)$\quad$ & \textbf{0.074 ± 0.018$^\dag$} & 0.29 ± 0.09 & \textbf{0.19 ± 0.12} & \textbf{0.18}$^{\star\star}$\\ 
\bottomrule
\end{tabular}
\begin{tablenotes}
\scriptsize{
\item$^{*}$ Significant difference compared to our proposed approach Att ($p < 0.05$, paired two-tailed t-test)
\item$^\dag$ Significant difference compared to fMRI-only Base model ($p < 0.05$, paired two-tailed t-test)
\item$^{\star}$ / $^{\star\star}$ Significant difference from 0 ($p < 0.05$ / $p < 0.01$, two-tailed t-test)}
\end{tablenotes}
\end{threeparttable}
\end{table}

As seen in the ablation study results, training the attention module with demographic or genetic data alone produces similar results to training with both modes of data together in our full model. All attention-based models performed significantly better than the unimodal fMRI model as measured by MSE. Using genetic data alone produced a slightly lower overall correlation. Although the performance levels are similar, including both genetic and demographic information may result in better understand of the interplay between these different types of individual features in the context of understanding ASD. 


\section{Conclusions}

In this work, we proposed an attention-based approach to integrating genetic and demographic information with fMRI data. The genetic and demographic data are used to guide attention to important fMRI encoded features. In a 10-fold cross-validation framework with a dataset of 228 ASD and TD subjects, We demonstrated improved performance in an ASD classification and ASD severity prediction task compared to other standard approaches of combining multimodal data. Future work will explore other genetic variables to use as input (e.g., number of genes contained in CNVs) and analyze the neuroimaging biomarkers that are used by the attention-based model for ASD classification and severity prediction. 

\subsubsection{Acknowledgements} This work is supported by ***.

\clearpage
%
%
%
%
\bibliographystyle{splncs04}
\bibliography{refs.bib}
\end{document}